\documentclass[letterpaper,twocolumn,fleqn]{article} 

\usepackage{ist}
\usepackage{siunitx}

\title{Realistic Image Degradation with Measured PSF}

\author{Christian Wittpahl, Hatem Ben Zakour, Matthias Lehmann, Alexander Braun}

\date{} 

\begin{document} 

\maketitle 

\thispagestyle{empty} 


\begin{abstract}
Training autonomous vehicles requires lots of driving sequences in all situations\cite{zhao2016}. Typically a simulation environment (software-in-the-loop, SiL) accompanies real-world test drives to systematically vary environmental parameters. A missing piece in the optical model of those SiL simulations is the sharpness, given in linear system theory by the point-spread function (PSF) of the optical system. We present a novel numerical model for the PSF of an optical system that can efficiently model both experimental measurements and lens design simulations of the PSF. The numerical basis for this model is a non-linear regression of the PSF with an artificial neural network (ANN). The novelty lies in the portability and the parameterization of this model, which allows to apply this model in basically any conceivable optical simulation scenario, e.g. inserting a measured lens into a computer game to train autonomous vehicles. We present a lens measurement series, yielding a numerical function for the PSF that depends only on the parameters defocus, field and azimuth. By convolving existing images and videos with this PSF model we apply the measured lens as a transfer function, therefore generating an image as if it were seen with the measured lens itself. Applications of this method are in any optical scenario, but we focus on the context of autonomous driving, where quality of the detection algorithms depends directly on the optical quality of the used camera system. With the parameterization of the optical model we present a method to validate the functional and safety limits of camera-based ADAS based on the real, measured lens actually used in the product. 
\end{abstract}

\section{Optical models for camera system validation}
\label{sec:intro}
For camera-based ADAS and for autonomous driving the functional and safety limits of the used camera systems need to be {\em quantitatively} determined\cite{Shashua2017}. Numerical test methods (SiL and HiL) play a central role in this functional validation. What is required is a process in which certain production parameters or tolerances are systematically varied over a given range - e.g. operating temperature range - and the optical quality of the camera system is evaluated. In a SiL or HiL setup the images and video sequences are then modified such that the camera looks as if it were operating at that set of parameters (e.g. at high temperature). Unfortunately, there currently exists no flexible optical model that allows this process in a comprehensive and efficient manner. 

In principle the optical properties of a lens are completely described by the Optical Transfer Function (OTF) in frequency space, or accordingly by the Point Spread Function (PSF) in image space \cite{Boreman2001}. The PSF -- with which we are concerned in this paper -- is a highly non-linear function with no apparent parameterization for mass production tolerances, and hence there exists no error model for mass production cameras, neither analytical nor numerical. 

At first glance this may seem strange, because the Zernike polynomials were expressedly developed to describe aberrations in optical systems \cite{Zernike1934}\cite{Born1999}. Zernike polynomials are successfully used to develop extremely complex optical system (prominent example: James Webb Space Telescope \cite{Choquet2014}). Nonetheless, the Zernike polynomials are ineffective for mass production lenses, as the aberrations are huge in comparison to diffraction-limited optical systems like telescopes or microscopes. In mass production lenses a large number of Zernike polynomials (50 - 100) need to be taken into account, rendering this process unusable for fitting or modeling of real, measured lenses. 

A lens is designed by help of a appropriate software like OpticStudio, Code V or OSLO. The PSF is a standard output of these software packages, and the SW can take tolerances and temperature shifts into account. But then the whole software package itself would become the optical transfer function, which is from a numerical point of view completely impractical for high-dimensional tolerance studies. Also, these packages lack a central requirement: sequential raytracing offers only limited to no possibility for depth in object space, which is mandatory for real 3d-scenes such as those used in SiL or HiL setups for the automobile industry. 

There are publications that use ANNs to estimate optical aberrations, cf. \cite{Weddell2007} or \cite{Schuler2013}. Both publications do not address the problem of parameterization for mass production. The first publication is very application specific to atmospheric phase-front aberrations for telescopes, and it is not obvious how to generalize the process to the SiL/HiL requirements presented here. The blind deconvolution of the second publication is even more limited, working on 2d images only (no depth information) and using a spatially invariant PSF, which is exactly not what we require. 

In summary there exists no numerical optical model for the PSF (or the OTF) to be used in the context of SiL/HiL validation of camera systems for the automotive industry.

\section{PSF modelling}
\begin{figure}
\centering
\includegraphics[width=\linewidth]{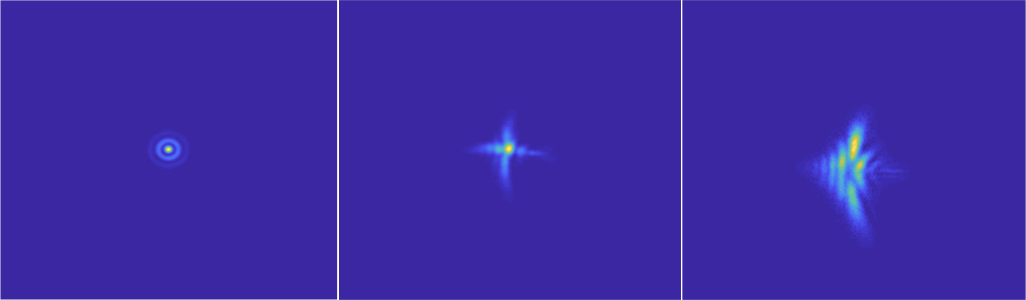}  
\\
(a) \hspace{0.3\linewidth} (b) \hspace{0.3\linewidth} (c)
\caption{Three example PSFs for different image field parameters $(\Delta z, R, \varphi)$. {\bf (a)} (0, 0, 0), {\bf (b)} (\SI{11.25}{\micro m}, \SI{2.25}{mm}, 0) and {\bf (c)} (0, \SI{3.00}{mm}, 0). Vertical displacement in (c) due to lens distortion. {\em Fehlt: Maßstab \SI{6}{\micro m}!}}
\label{fig:PSFOverview}
\end{figure} 
The PSF is a highly non-linear function that represents the transfer function of an optical system. A typical real automotive lens has 4 to 6 lens elements, possibly an IR-cut filter included as seventh element, and therefore some 8 to 14 optically active surfaces (some elements may be cemented in subgroups). Each surface has its own set of tolerances, which are at least 3 position and 3 direction variables - excluding the topic of surface deviations. Rotational symmetry can apparently not be assumed. Fig.~\ref{fig:PSFOverview} qualitatively depicts three measured PSFs from the same lens at different image field, and clearly demonstrates the strong variation in the quality of the intensity distribution. What is required is a mathematical-numerical model that describes the properties of real lenses as a function of a limited set of dependent variables.

\subsection{Non-linear regression with ANN}
\label{sec:ann_regression}
Regression and especially non-linear regression with artificial neural networks for highly asymmetric functions is an established and ongoing field of research (e.g.  \cite{Rosenblatt1962}\cite{Rumelhart1986}\cite{Belagiannis2015}). The ANN 'learns' the function by training it with an appropriate number of examples. During training both the function training set as well as the according parameters are used as input to the first layer of the ANN. During operation only the input parameters are used, and the ANN evaluates the function at the given parameter position. 
\begin{figure}
\centering
\includegraphics[width=0.7\linewidth]{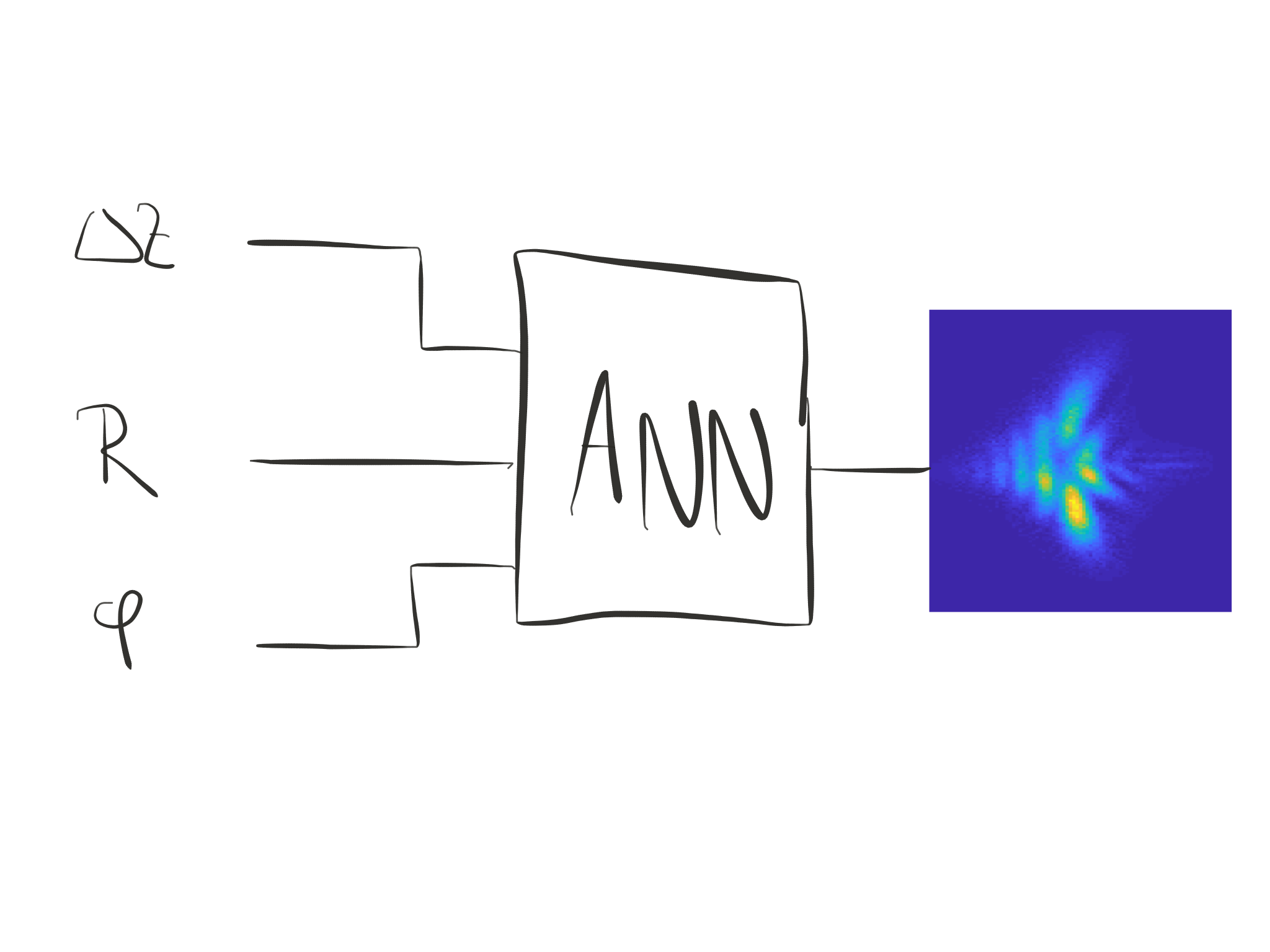}
\caption{Model overview: the ANN takes a number of input parameters (in principle variable) and outputs a PSF. In this report the model parameters are defocus $\Delta z$, image height $R$ and azimuth $\varphi$, according to the measurement parameters.} 
\label{fig:modelOverview}
\end{figure} 
The goal of our novel model is depicted in Fig.~\ref{fig:modelOverview}. The ANN takes a (limited) number of input parameters and evaluates the output PSF as a function of those parameters. The input parameters in principle are variable and depend on the actual simulation goal. E.g. in a tolerance calculation some tolerance measure might be used as input into the ANN. In this report we use three input parameters to the ANN, defocus $\Delta z$, image height $R$ and azimuth $\varphi$. These parameters accord to the three measurement parameters (see next Sec. Measurement). But the strength of our approach is that it is very flexible in the number and type of inputs, such that it can be used in a broad variety of different simulation scenarios. We have used mean square error as the distance metric to train the ANN. Details of this training process will be published separately.

\subsection{Measurement}
\label{sec:PSFMeasurement}
\begin{figure}
\centering
\includegraphics[width=0.8\linewidth]{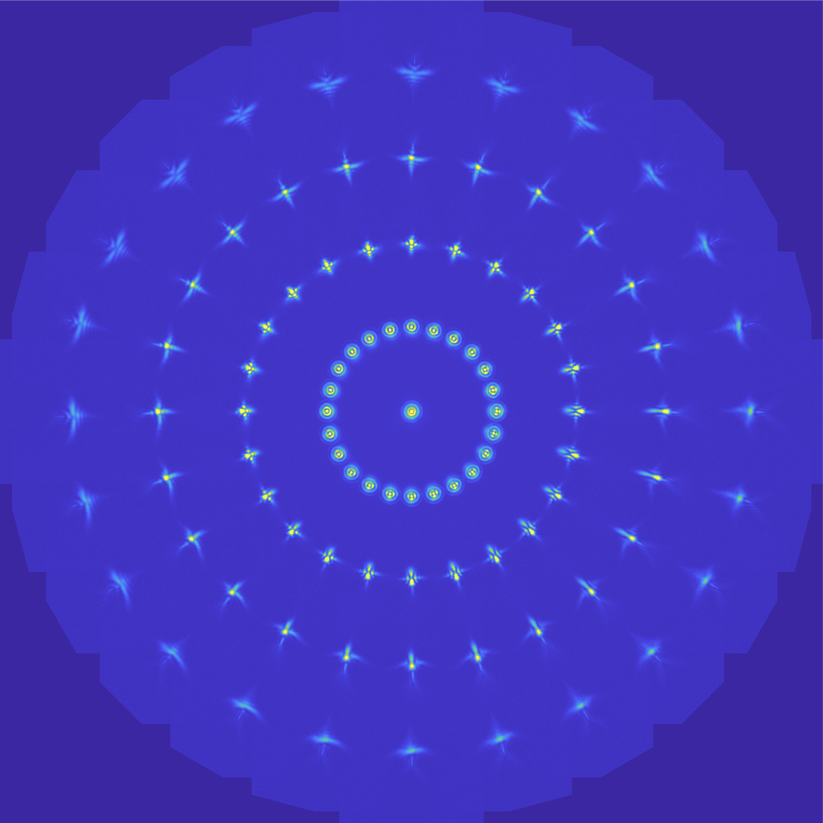}
\caption{Measurement overview with 108 measured PSFs for a single defocus value $\Delta z = \SI{0}{\micro m}$. Radius varied from $R = \SI{-3,00}{\milli\metre}$ to $R = \SI{+3,00}{\milli\metre}$, azimuth varied from \SI{0}{\degree} to \SI{165}{\degree}. Position to scale, PSF enlarged for visibility.} 
\label{fig:MeasurementOverview}
\end{figure} 
The measurement data were taken in collaboration with the company trioptics in Wedel, Germany. The used measurement system was a Trioptics ImageMaster HR. The used setup has an effective pixelsize of $d_\mathrm{pixel, effective} = \SI{0,3070}{\micro\metre}$. Due to the combination of the photopic vision filter and a monochromatic CCD-sensor the measured data does not cover chromatic aberrations. This is left for future work. Overall 27 lenses were measured, based on three different lens designs for automotive camera modules. For this work we selected just a single lens, meaning that the training of the neural network is based solely on the measurement data of one lens specimen. The lens has a focal length of \SI{6}{mm} and a field-of-view of \SI{60}{\degree}. 

The measurement has three parameters: defocus $\Delta z$, image height $R$ and azimuth $\varphi$. The image height was varied from $R = \SI{-3,00}{\milli\metre}$ to $R = \SI{+3,00}{\milli\metre}$, azimuth full circle, and defocus from \SI{-50,0}{\micro m} to \SI{+50,0}{\micro m}. Due to measurement time restrictions the parameter sampling is not evenly distributed. Basically two measurement series per lens were recorded: one with high in-plane resolution for $R$ and $\varphi$ and low defocus resolution $\Delta z$, and one with reduced in-plane resolution and high resolution and larger range for the defocus. The first series resulted in 243 measured PSFs, the second in 972 PSFs. Fig.~\ref{fig:MeasurementOverview} shows the measured PSFs for the used lens in a single plane of defocus, i.e. all the shown PSFs have the same defocus value of $\Delta z = \SI{0}{\micro m}$. Note that the PSFs positions are to scale in Fig.~\ref{fig:MeasurementOverview}, but the actual PSFs have been enlarged to make them visible. Therefore, the outer circle accords to the image height $R = \SI{3,00}{\milli\metre}$, but the real PSFs are much smaller. 

\subsection{Downsampling}
The first important step is to downsample the high resolution scans of the PSF: first, if the target image sensor has a pixel size of (e.g.) \SI{3}{\micro m} the high resolution of the measurement (\SI{0,3070}{\micro\metre}) is not necessary. Second, the resolution determines the size of the ANN, the amount of training data and hence the required computing resources. For this article we cropped and downsampled the data to a pixel size of approximately \SI{6}{\micro m}, resulting in a work resolution of 13x13 for the PSF, mainly restricted due to the limited computing power. An example of this process is shown in Fig.~\ref{fig:modelOverview}. Based on this input data we have varied both the structure of the ANN and the used learning algorithm. 
\begin{figure}
\centering
\includegraphics[width=\linewidth]{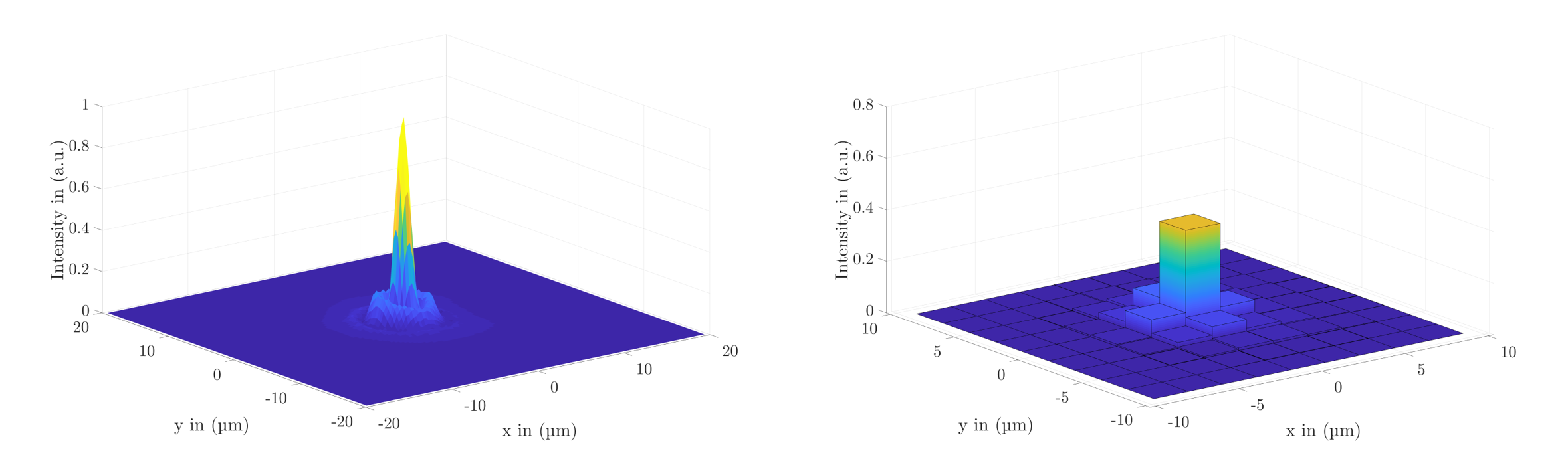}
\caption{The original measurement data (left) is cropped and downsampled to approximate real pixel size (right) and in order to reduce the size of the ANN.} 
\label{fig:modelOverview}
\end{figure}

\subsection{PSF model example}
\begin{figure}
\centering
\begin{tabular}{c}
\includegraphics[width=\linewidth]{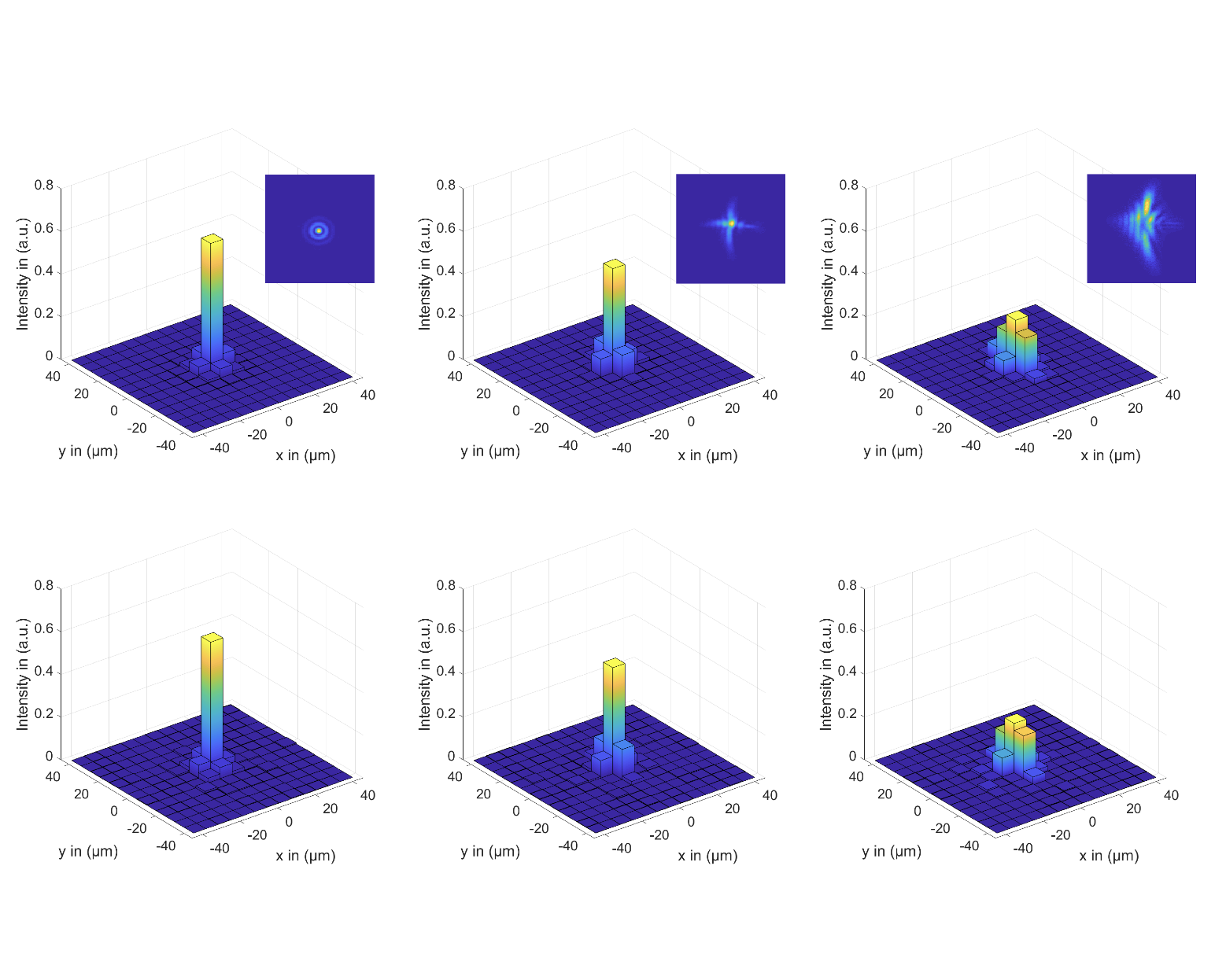}
\end{tabular}
\caption{Three exemplary modelling predictions for different field situations (cf. text, same as Fig.~\ref{fig:PSFOverview}). Upper row: measurement. Lower row: model prediction. Inset: high resolution PSF.} 
\label{fig:predictionExample}
\end{figure} 
Fig.~\ref{fig:predictionExample} shows an example of the prediction process for a given lens sample.  The three columns a, b and c represent three different field positions $(\Delta z, R, \varphi)$, the same as in Fig.~\ref{fig:PSFOverview}: (a) (0, 0, 0),  (b) (\SI{11.25}{\micro m}, \SI{2.25}{mm}, 0) and  (c) (0, \SI{3.00}{mm}, 0). The upper row shows the actual measurement, the lower row the output from the ANN for the given input parameters. The inset displays the high resolution measurement image of the respective PSF. The MSE distance metric yields satisfactory agreement between the measurement and the model. Improving on the accuracy of the prediction will be a focus of our further studies. Nonetheless, the ANN can model the distinctly varying spatial distribution of the PSF, especially the changing quality is modeled faithfully. In summary we therefore can apply the model PSFs as a transfer function to existing images. 

\section{Implementation}
\label{sec:implementation}
The goal of a validating SiL or HiL setup is to simulate or modify scenarios to look like they were taken under different circumstances. For example, the temperature expansion of the plastic holder may lead to a defocus $\Delta z$, giving a blurred image. Therefore the image data -- either simulated or already recorded data -- needs to be modified to reflect these circumstances. For the optical question at hand this means that the existing images need to be convolved with the appropriate PSF as a transfer function. This section describes the steps we have used to convolve existing images efficiently with the predicted PSF from our ANN. 

\begin{figure}
	\centering
	\begin{tabular}{c}
		\includegraphics[width=\linewidth]{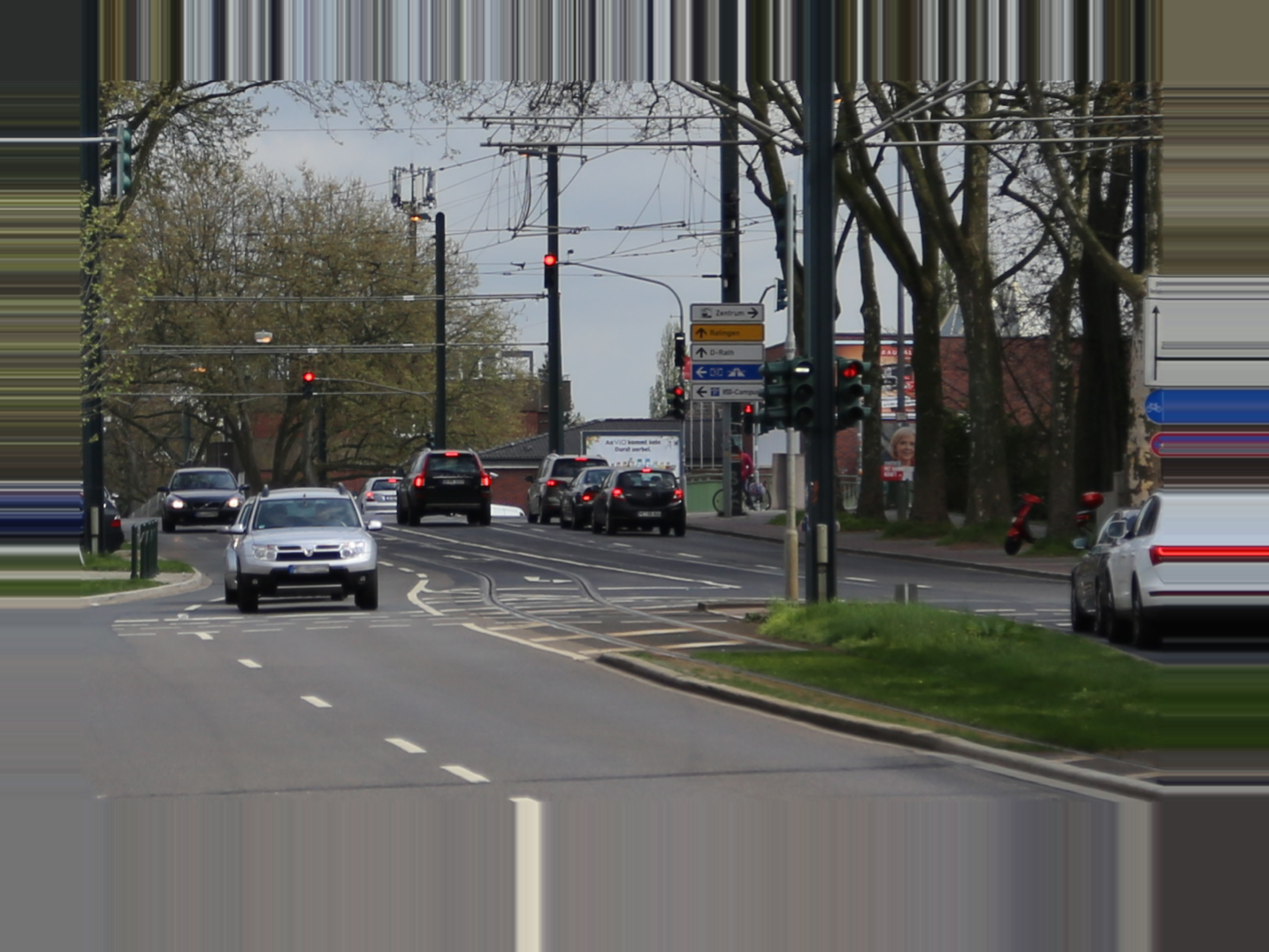}
	\end{tabular}
	\caption{Example Image taken in D\"usseldorf, with constant value border continuation.} 
	\label{fig:exampleMuenster}
\end{figure}
\begin{figure}
	\centering
	\begin{tabular}{c}
		\includegraphics[width=\linewidth]{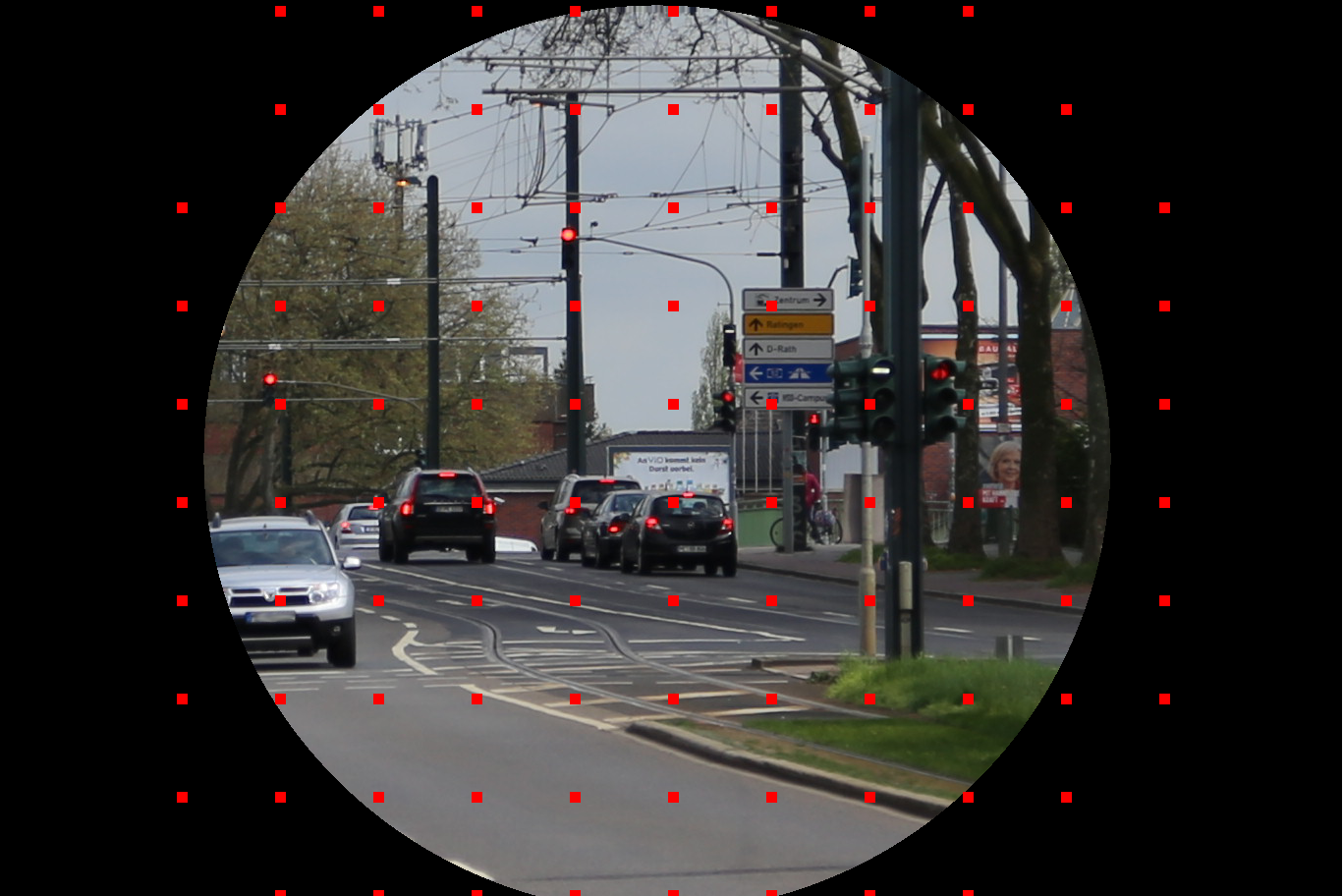}
	\end{tabular}
	\caption{Example Image with valid image area and sampled points} 
	\label{fig:exampleMuensterValid}
\end{figure}

 
The filter process will be described using the example image in Fig.~\ref{fig:exampleMuenster}. The image was taken with a high quality consumer camera (Canon..., sensor resolution w x h, pixel size ...). Since our PSF measurements are only valid up to a certain image height, we black out every pixel outside of this range, imitating an undersized aperture on the image sensor with given pixel size. The aperture is visible in Fig.~\ref{fig:exampleMuensterValid}.

Using our ANN model we can now map a unique PSF to each valid pixel in any image of a sensor. Optically this would accord to the real physical situation that the light from every viewing direction within the angular resolution (which following Fourier optics is of course determined by the pixel size) travels through the lens on its own path, and hence has its own PSF. Convolving every single pixel with its appropriate PSF is numerically cumbersome though, and we opted for a more practical approach by manually determining ROIs and then interpolating the results. The red dots in Fig.~\ref{fig:exampleMuensterValid} represent the corners of ROIs. It actually would be an interesting research question in its own right to determine a quantitative limit for the size of this ROI, or if a separate PSF for every pixel is required. Note that every pixel of the target image is of course convolved with a PSF, but neighbouring pixels are treated with (almost) the same PSF, and the exact PSF is determined by bilinear interpolation as will be detailed in the following section. 

\subsection{Interpolation}
\begin{figure}
	\centering
		\includegraphics[width=\linewidth]{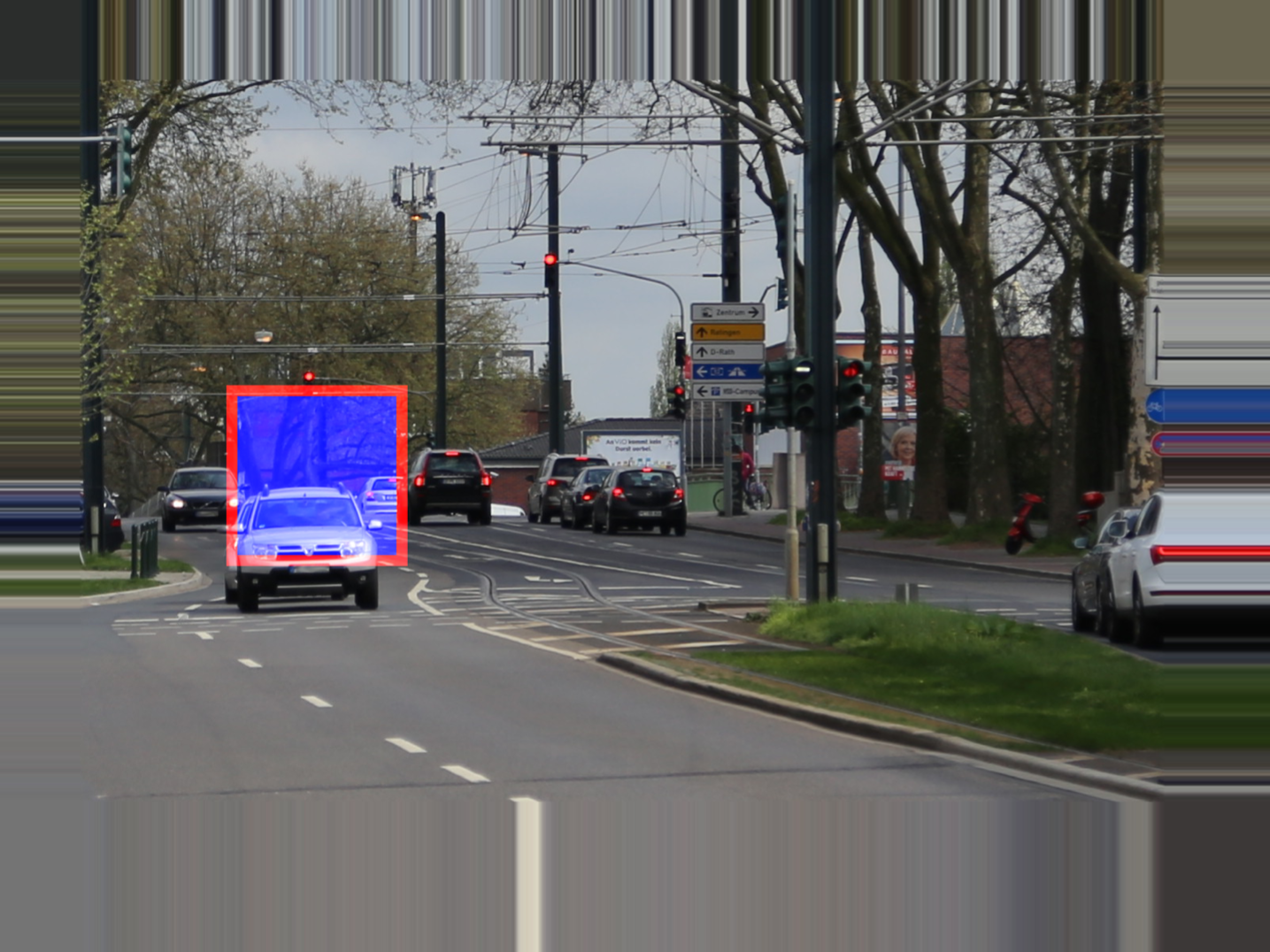}
	\caption{Example of a rectangular area for convolution.} 
	\label{fig:exampleMuensterBorderBlock}
\end{figure} 
\begin{figure}
\centering	
\includegraphics[width=\linewidth]{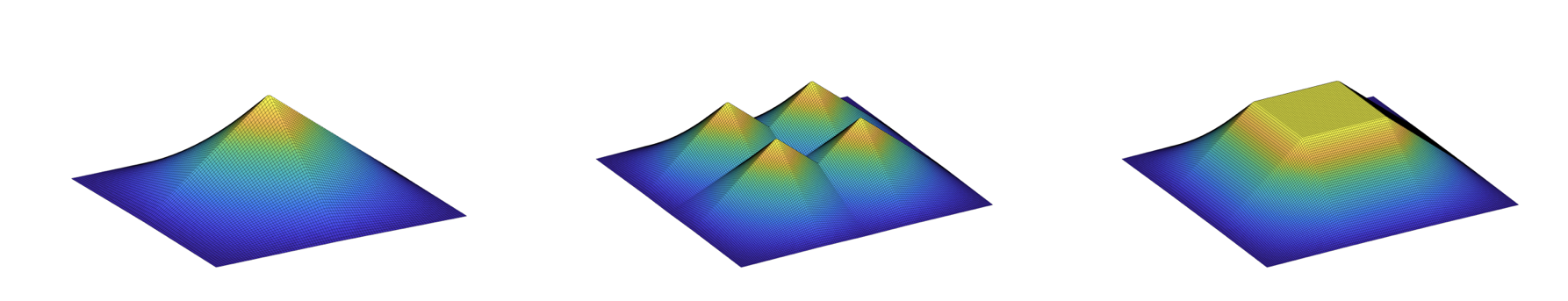}
\caption{Construction of the weighting function for the bilinear interpolation of the PSF. Left: single kernel for simple bilinear weighing, middle: all four kernels, right: resulting final kernel.} 
\label{fig:bilinearInterpolation}
\end{figure} 
\begin{figure}
	\includegraphics[width=\linewidth]{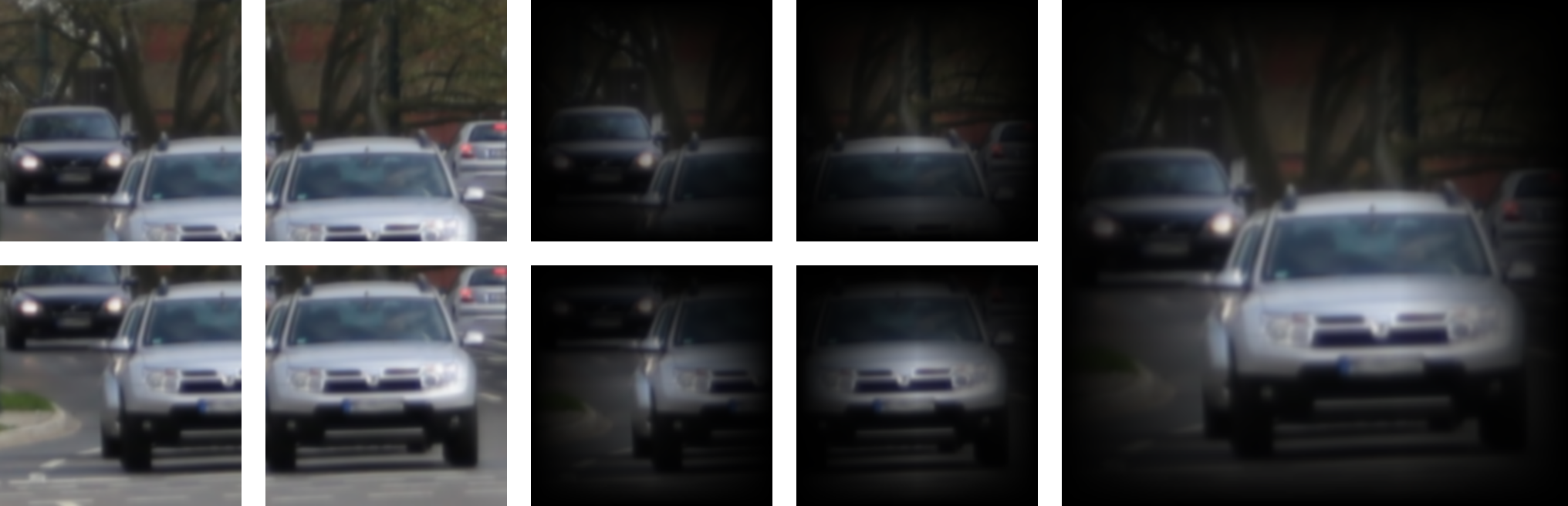}
	\caption{Sum of weighted areas. Left: cropped image parts, middle: weighted image parts, right: resulting weighted and convolved degraded image part.}
	\label{fig:exampleSummedAreas}
\end{figure}
Since kernels of pixels which are positioned closely together are similar, we can sample a smaller number of PSFs and interpolate the fully convoluted image based on those. To achieve this we choose a uniformly spaced grid of pixels across the image (Fig.~\ref{fig:exampleMuensterValid}). 

Every valid pixel of the image will be inside a rectangle consisting of four points in this grid. Using bilinear interpolation, we can calculate a new PSF kernel based on those. The maximum error of this interpolation can be estimated by comparing the bilinear interpolation's result to the ANN's. 
Every sampled point is the corner of four rectangular parts of the image. These four areas create a new ROI with the sampled point at its center as shown in Fig.~\ref{fig:bilinearInterpolation}.

Weighting four of these bigger rectangles appropriately and adding their values we get the exact result of a convolution with the corresponding PSFs, that could have been calculated through bilinear interpolation. The weighting scheme and how four of these weighted areas sum to a whole are visible in Fig.~\ref{fig:bilinearInterpolation}.

Fig.~\ref{fig:exampleSummedAreas} shows the application of this weight function to the ROI from Fig.~\ref{fig:exampleMuensterBorderBlock}. The right image thus is the result of a convolution of the whole area with four different PSFs only, instead of the 160 different PSFs according to a pixel-exact model.

\subsection{Degraded image}
\begin{figure}
	\includegraphics[width=\linewidth]{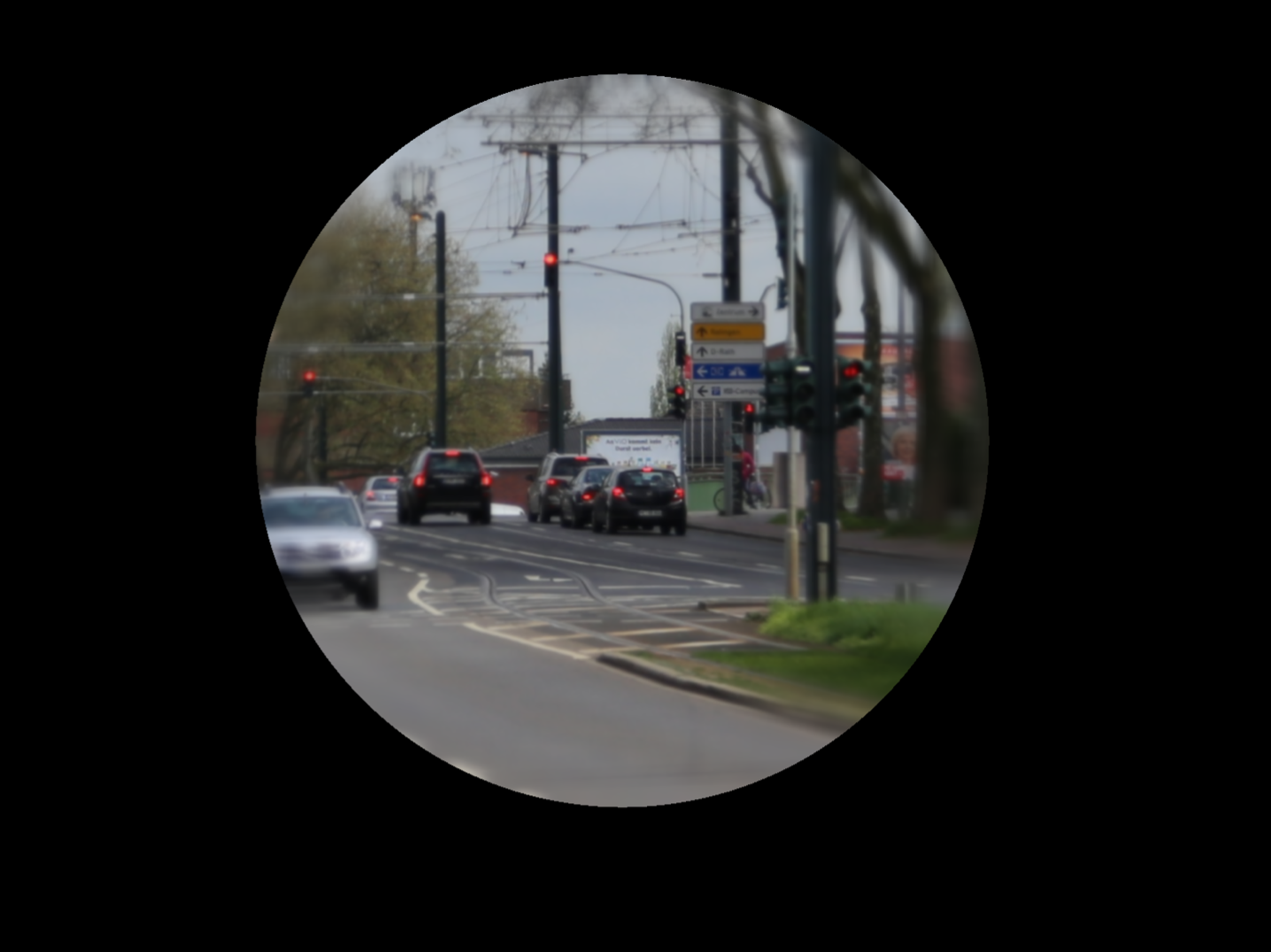}
	\caption{Resulting degraded image after applying the spatially variant PSF, with a defocus value of $\Delta z = 0$.}
	\label{fig:result}
\end{figure}

The resulting degraded image of this process is depicted in Fig.~\ref{fig:result}. There the spatially varying PSF is convolved with the original image with the right PSF at the appropriate position. For completeness we also applied a standard blind deconvolution to minimize the influence of the original lens (which probably wasn't necessary considering the huge difference in optical quality between a state-of-the-art consumer lens and a series production automotive lens, but was also included in the algorithm as a place holder function for a more rigorous examination). Because the physical size of the original image is larger then the maximum radius of our measurement ($R_\text{max} = \SI{3}{mm}$) the performance of the measured lens strongly declines toward the edge of the image circle. In a real application the size of the imager would be selected as a rectangle distinctly within the circle, hence the very strong blurring at the edges would not be visible. 

Note that for this simulation the defocus value is $\Delta z = 0$ for the whole image, i.e. the depth information (or an inclusion of the field curvature) is not present. In that sense Fig.~\ref{fig:result} is similar to the output of a standard lens design software where the user chooses a function to trace an example image through the lens design. What differentiates the optical model presented in this paper is the ability to include depth information in the formation of the resulting image. Therefore the next section discusses depth. 

\section{Depth}
\label{sec:depth}
Camera-based ADAS are fixed-focus systems, where the lens-imager position is determined once and for all with the alignment procedure during production. Therefore objects with different distances (in object space) will also have different image distances in relation to the principle planes, i.e. they will have a defocus value $\Delta z$ with respect to the image plane that was fixed during alignment (which in real lenses is of course not a plane but the field curvature). As mentioned in the introduction the strength of this optical model is that it allows for this depth information and hence the exact focus position (or rather the defocus value $\Delta z$) to be used, which is -- to the best of our knowledge -- a new feature for optical simulations. 

This idea is applicable in both optical ray tracing and computer game like z-Buffering. There are different possibilities to calculate the value for $\Delta z$ in these scenarios. As a simple example consider an OpenGL-like 3d-engine that uses a z-buffer, which is just a measure for the depth information (hence the name $z$). The z-buffer information is linearized and scaled to yield real $z$ values (in object space), and using the simple lens formula $1/f = 1/o + 1/i$ (with $f$ focal length, $o$ object distance and $i$ image distance) a defocus value $\Delta z$ in relation to a given image plane can be determined. 

\begin{figure}
	\includegraphics[width=\linewidth]{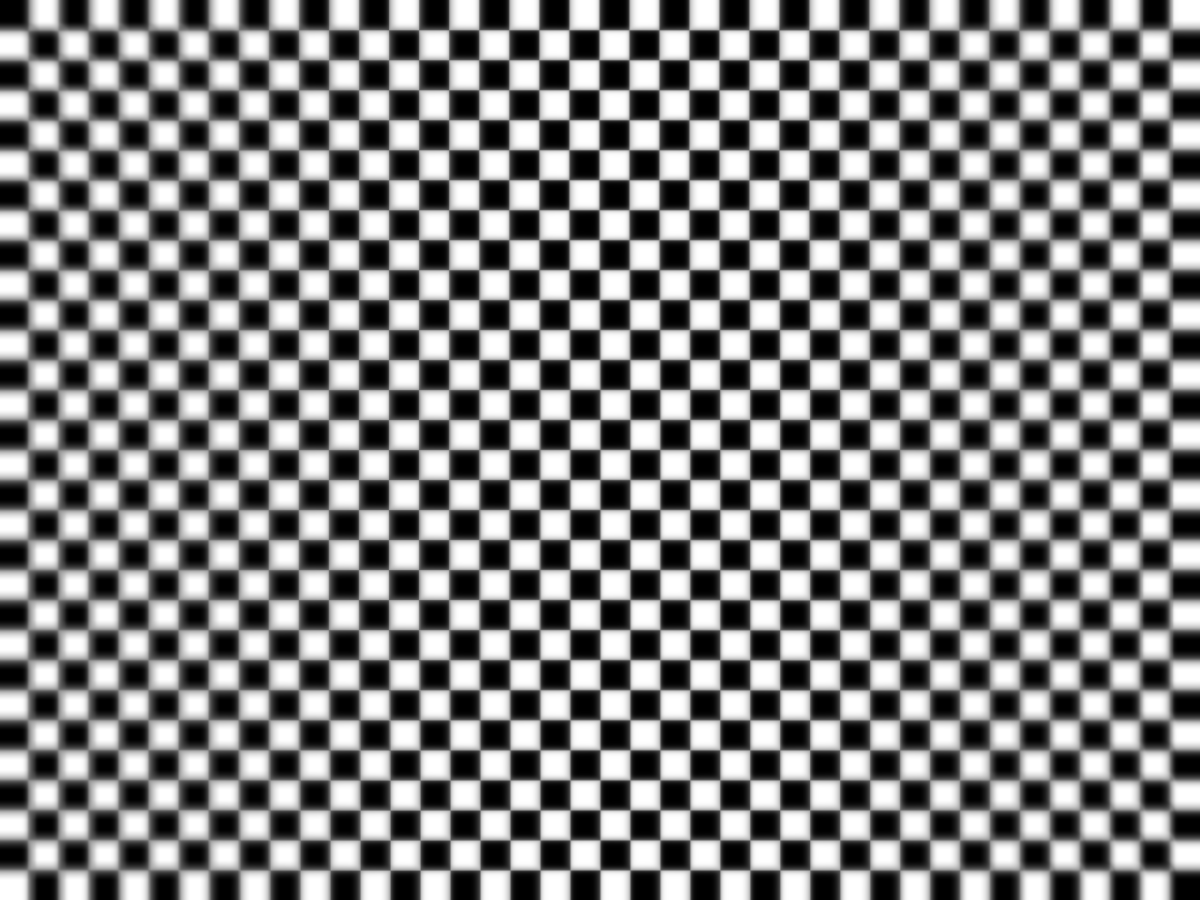}
	\caption{Linear depth gradient applied to a checkerboard. Please note the slight differences between the left and right edge of the image, resulting from positive and negative $\Delta z$ values.}
	\label{fig:depthCheckerboard}	
\end{figure}
To demonstrate this feature we have applied the model with a linear depth gradient to a simple checkerboard (see Fig.~\ref{fig:depthCheckerboard}). In the image the value for $\Delta z$ is varied from left to right from positive maximum measurement value (here: \SI{+50}{\micro m}) to negative maximum (\SI{-50}{\micro m}).  This setup corresponds roughly to a checkerboard in object space that is strongly tilted with respect to the optical axis of the camera system. 

For clarity we used a spatially invariant PSF, i.e. we did not use the position information $R$ and $\varphi$. Thus the blurring stems only from the value of the defous $\Delta z$. This is of course unrealistic, but this way the effects of depth and decreasing lens performance with large field are separated. Looking closely at the left and right edge of Fig.~\ref{fig:depthCheckerboard} there are subtle differences in the blurring, because the lens exhibits an asymmetric PSF function for positive and negative values of $\Delta z$. This shows nicely the new quality of simulation possible with our novel optical model. 

\section{Summary and outlook}
With the presented universal optical model we are now able to included measured lens data within optical simulations, for both ray tracing and z-buffering. For the context of testing and validating algorithms for autonomous driving this opens a new quality currently not found in SiL or HiL test setups. As an example, the thermal expansion of a camera head is a central validation question for every camera maker, mainly the defocussing effect of the thermal expansion of the plastic holding components. This defocussing is expressed as $\Delta z$. Therefore with the new model you can now setup a SiL or HiL simulator that examines the camera response of your real algorithm to the blurring effect or your real measured PSF. Systematically varying $\Delta z$ then allows you to systematically test and validate the function and safety limits of your own ADAS algorithms with respect to the temperature expansion of the camera head. 

This work shows the feasibility of our approach. There are several different optimizations and expansions we are pursuing. First of all is the spatial resolution, both of the measurement (how many sampling points to you need for a faithful model?) as well as the model itself (downsampling resolution? One PSF for every pixel in the target image?). The distance metric for the training of the neural network can also be varied to reflect spatial information, like CNNs. Finally, the direct comparison of a simulated driving scene and its real, measured counterpart will be an important step in demonstrating the ability and the limits of our new universal optical model. 


\clearpage
\small
\bibliographystyle{unsrt}
\bibliography{Literature}






\begin{biography}
Christian Wittpahl received his B.Eng. from the University of Applied Sciences in D\"usseldorf in 2017. His thesis work forms the basis for this publication, where he developed the degradation algorithms, with a focus on efficient convolution. Currently he is a master student at the same University, studying electrical engineering. 

Hatem Ben Zakour received his B.Eng. from the University of Applied Sciences in D\"usseldorf in 2017. In his thesis work he implemented the first neural networks for the regression of this publication. Currently he is a master student at the same University, studying electrical engineering. 

Matthias Lehmann received his B.Eng. in 2015 and his M.Sc. in 2017 from the University of Applied Sciences in D\"usseldorf. His bachelors thesis examined the influence of noise on MTF algorithms and measurements. In his master thesis he developed the SiL environment used for this publication. He is currently pursuing his PhD with a focus on the presented universal optical model, and its application to test and validate algorithms for autonomous driving. 

Alexander Braun received his diploma in physics with a focus on laser fluorescent spectroscopy from the University of G\"ottingen in 2001. His PhD research in quantum optics was carried out at the University of Hamburg, resulting in a Doctorate from the University of Siegen in 2007. He started working as an optical designer for camera-based ADAS with the company Kostal, and became a Professor of Physics at the University of Applied Sciences in D\"usseldorf in 2013, where he now researches  optical metrology and optical models for simulation in the context of autonomous driving. He's member of DPG and SPIE, participating in norming efforts at IEEE (P2020) and VDI (FA 8.13), and currently serves on the advisory board for the AutoSens conference. 

\end{biography}

\end{document}